\documentclass[12pt]{article}
\usepackage{setspace}
\usepackage{caption}
\usepackage{amsfonts}
\usepackage{graphicx}
\usepackage{color}
\usepackage{amsmath}
\usepackage{float}
\usepackage{comment}
\usepackage{epsf}
\usepackage{hyperref}
\usepackage{epstopdf}
\newcommand {\be}{\begin{equation}}
 \newcommand {\ee}{\end{equation}}
 \newcommand {\bea}{\begin{array}}
 
 \newcommand {\eea}{\end{array}}

\textwidth=6.5in \hoffset=-.75in \voffset=-1in \numberwithin{equation}{section}
\numberwithin{figure}{section}

\begin{document}

\begin{titlepage}
%\bigskip \begin{flushright}
%\\
%\end{flushright}
%\maketitle
\vspace{1cm}
\begin{center}
{\Large \bf {Destroying a near-extremal Kerr-Newman-AdS black hole}}\\
\end{center}
\vspace{0.5cm}
\begin{center}
\renewcommand{\thefootnote}{\fnsymbol{footnote}}
Yu-Song{\footnote{songyu2011@163.com}}
\\
Institute of Modern Physics, Northwest University,Xi'an 710069, China\\
\vspace{0.5cm}
Rui-Hong Yue{\footnote{rhyue@yzu.edu.cn}}
\\
Center for Gravitation and Cosmology, College of Physical Science and Technology,\\
Yangzhou University, Yangzhou, 225009, China

\end{center}

\begin{abstract}
If radiative and self-force effects are neglected, we find that feeding a test particle into a near-extremal Kerr-Newman-AdS black hole could lead to destroy their event horizon, giving rise to naked singularities. Hence radiative and self-force effects must be taken into account to further test cosmic censorship. Moreover, the allowed parameter range for this test particle is very narrow, this leaves the possibility of radiative and self-force effects considerations to cure the problem of WCC violation in Kerr-Newman-AdS space time.
\end{abstract}
\end{titlepage}\onecolumn
\bigskip

 \section{Introduction}
\label{sec:intro}

According to the so-called cosmic censorship conjecture \cite{Penrose},
all physical singularities due to gravitational collapse should be hidden behind an event horizon.
This hypothesis that no naked singularity occurs in our universe can be described
in weak and strong versions \cite{Joshi:2008zz}. One way of testing the week
cosmic censorship conjecture (WCCC) is to throw a test particle into an existing black hole.
If the particle can pass through the horizon and change the parameters of the black hole
so that the horizon disappears, then the cosmic censorship conjecture might break down.

The seminal work was done by Wald \cite{wald}, which showed that a test particle
can not destroy the horizon of an extremal Kerr-Newman black hole. From then on, many authors followed the spirit of Wald's gadanken experiment and revisited the problem of WCC violation. Some of the
attempts involving test particles have been successful
to produce naked singularities \cite{Jacobson:2009kt,hubeny:1999,Richartz:2008,matsas:2009,Richartz:2011}. For example,
Jacobson et al. \cite{Jacobson:2009kt}  asserted that it would be possible to destroy a
Kerr black hole with infalling test particle if we start with a near-extremal
configuration, rather than an extremal black hole.
In \cite{hubeny:1999}, Hubeny showed that overcharging a near-extremal Reissner-Nordstrom(RN) black hole is possible by injecting a test particle into the black hole. For charged and spinning black holes, it is possible to destroy an extremal \cite{Gao:2012ca} and near-extremal \cite{Saa:2011wq} Kerr-Newman black hole by a test particle with electric charge and angular momentum by neglecting the backreaction effect.  The issue about the problem of cosmic censorship has been extensively studied in recent years \cite{hod}-\cite{Semiz2}.

Since the discovery of the AdS/CFT correspondence \cite{Maldacena:1997re,Gubser:1998bc,Witten:1998qj}, its developments have brought about many new insights into the nature of black
holes. The study of asymptotically anti-de Sitter black holes has
become more important and realistic than ever.
It is interesting to know the role of $\Lambda$ in the discussing turning a black hole into
a naked singularity by plugging a test particle into black hole.
Recently, Zhang et al \cite{Zhang:2013tba} asserted that an extremal Reissner-Nordstr$\ddot{o}$m
anti-de Sitter black hole can be overcharged by a test particle and both extremal Kerr-de-Sitter/anti-de-Sitter black holes can be overspun by a test particle, which implies a possible breakdown of the cosmic censorship conjecture. Moreover, we have also
recovered that a test particle may destroy the horizon of an extremal Kerr-Newman-AdS Black Holes
black hole, resulting in an apparent violation of the cosmic censorship \cite{songyu:2017}.
In this paper, we further study the near-extremal Kerr-Newman-AdS black holes, and will find that the possibility to destroy a near-extremal Kerr-Newman-AdS by plunging test particles across their event horizon when
neglecting the self-force, self-energy and radiative effects. We also provide a numerical plot showing that a test particle, which potentially could destroy the black hole, could really fall all the way from infinity into the black hole.

In Sec. \ref{sec.1} we show how a near-extremal Kerr-Newman-AdS black hole can be destroyed by plunging test particle across its event horizon. Finally in Sec. \ref{sec.2}, we give a closing remarks. In this paper we use the unit system where $c = G = \hbar = 1$.

\section{destroying Near-extremal Kerr-Newman-AdS black hole}\label{sec.1}

In four-dimensional spacetime, the Kerr-Newman-AdS metric takes the following
form in Boyer-Lindquist-type coordinates~\cite{songyu:2017, Aliev}

\begin{equation}\label{metric}
ds^2=-\frac{\Delta_r}{\rho^2} \left( dt - \frac{a\sin^2\theta}{\Xi} d\phi \right)^2
+\frac{\rho^2}{\Delta_r} dr^2 + \frac{\rho^2}{\Delta_\theta} d\theta^2+\frac{\Delta_\theta}{\rho^2}\sin^2\theta \left(adt-\frac{r^2+a^2}{\Xi} d\phi \right)^2,
\end{equation}
where
\begin{eqnarray}
&&\Delta_\theta=1-\frac{a^2}{l^2} \cos^2\theta,\quad
\rho^2 = r^2 + a^2 \cos^2\theta,\quad
\Xi=1-\frac{a^2}{l^2}\label{4kads},\nonumber\\
&&\Delta_r = \left(r^2 + a^2\right)\left(1 +\frac{r^2}{l^2}\right) -
2 M r + Q^2,\label{deltatheta}
\end{eqnarray}
in which $M$, $a$ and $Q$ is the mass, angular momentum per unit mass and charge of black hole, respectively.
$l$ is the curvature radius determined by the negative cosmological constant $\Lambda=-3 l^{-2}$.
The radius of the black hole's outer (event) horizon
and inner (Cauchy) horizon are given by \cite{Aliev}
\begin{eqnarray}
r_{+}&=& \frac{1}{2}\left(X+Y \right)\,\,,~~~~~~~r_{-}=
\frac{1}{2}\left(X-Y \right)\label{horizons}
\end{eqnarray}
respectively, where
\begin{eqnarray}
X&=&\sqrt{u-l^2-a^2}\,\,,~~~~~ Y= \sqrt{-u -l^2-a^2 +\frac{4 M
l^2}{X}}, \label{XY}\\
u&=&\,\frac{l^2+a^2}{3}+\frac{l^{\,4/3}\,\left(M_{1e}^2-M_{2e}^2
\right)^{2/3}}{\left(2 N^2- M_{1e}^2-M_{2e}^2 \right)^{1/3}} +
l^{\,4/3} \left(2 N^2- M_{1e}^2-M_{2e}^2 \right)^{1/3},\label{newu}\\
N^2&=&M^2+\sqrt{\left(M^2-M_{1e}^2\right)\left(M^2-M_{2e}^2\right)},\nonumber
\end{eqnarray}
and
\begin{eqnarray}
M_{1e}=(l/\sqrt{54})\,\sqrt{\zeta + \eta^3},\\
M_{2e}=(l/\sqrt{54})\, \sqrt{\zeta -\eta^3}.
\end{eqnarray}
Here
\begin{eqnarray}
\label{zeta} \zeta = \left(1+\frac{a^2}{l^2}\right)\left[\frac{36
\left(a^2+Q^2 \right)}{l^2}-\left(1+\frac{a^2}{l^2}\right)^2
\right],\quad
\eta = \left[\left(1+\frac{a^2}{l^2}\right)^2+ \frac{12
\left(a^2+Q^2\right)}{l^2}\right]^{1/2},
\end{eqnarray}
where $M_{1e}$ and $M_{2e}$ is two extremal mass parameters, it is easy to show that only the mass parameter $ M_{1e} $ has a
definite physical meaning, the black hole mass parameter $M$ must satisfy the relation\cite{Aliev}
\begin{equation}
M\geq M_{1e}.\label{extremalmass},
\end{equation}
The black hole become extremal when $M_{1e} = M$, which leads to
\begin{equation} \label{extremalcond}
u_{ex}=\frac{a^2+l^2}{3}+\frac{2}{3} l^2 \sqrt{\left( 1 + \frac{a^2}{l^2}\right)^2 + \frac{12
\left(a^2 + Q^2\right)}{l^2}}
\end{equation}
From Eqs.(\ref{horizons}),(\ref{XY}), one finds that the Kerr-Newman-AdS black hole become extremal
when
\begin{equation} \label{extremalcondu}
Y= -u_{ex} -l^2-a^2 +\frac{4 M l^2}{X}=0,
\end{equation}
where the inner and outer horizons coincide with each other.
We call the near-extremal Kerr-Newman-AdS black hole for which
\begin{equation} \label{nearextrformalcon}
-u_{ex} -l^2 - a^2 +\frac{4 M l^2}{X} = \delta
\end{equation}
with $0 << \delta\ll M$. Then we can obtain
\begin{eqnarray}
&&M = M_{ne}\nonumber\\
&& = \frac{\sqrt{-a^2-l^2+\frac{1}{3}l^2\sqrt{\left(1+\frac{a^2}{l^2}\right)^2+\frac{12
\left(a^2+Q^2\right)}{l^2}}} [a^2+l^2+\delta+\frac{1}{3}l^2 \sqrt{\left(1+\frac{a^2}{l^2}\right)^2+\frac{12
\left(a^2+Q^2\right)}{l^2}}]}{4l^2}.
\label{absorbcon}
\end{eqnarray}
We consider a test particle with electric charge $q$, energy $E$, and orbital angular $L$ captured by a near-extremal Kerr-Newman-AdS black hole.
After capturing the test particle, the black hole is described by the changed parameters. \cite{Misner}
\begin{eqnarray}
a'=\frac{a M+L}{M+E}\,,
M'=M+E\,,
Q'=Q+q\,,\label{newpara}
\end{eqnarray}
where $'$ is used to describe
the changed parameters after capturing a test particle. Then the new extremal condition can be written formally as
\begin{equation} \label{afterextrema}
-u' -l^2-a'^2 +\frac{4 M'_{ne} l^2}{X'} = 0.
\end{equation}
If a Kerr-Newman-AdS black hole turns into a naked singularity, the following condition must be satisfied
\begin{eqnarray}
-u' -l^2-a'^2 +\frac{4 M'_{ne} l^2}{X'} < 0,\\
X' > 0 . \label{nearconhditionX}
\end{eqnarray}
For simplicity, the above condition can be defined formally as
\begin{eqnarray}
F'(a',M'_{ne},Q',l) = -u' -l^2-a'^2 +\frac{4 M'_{ne} l^2}{X'} = F(a,M,Q,l,\delta,E,L,q) < 0,\label{aftercaptureNear}\\
X'(a',M'_{ne},Q',l) = X(a,M,Q,l,\delta,E,L,q) > 0 .\label{afterX}
\end{eqnarray}
Eq.(\ref{aftercaptureNear}) means that the event horizon of black hole is destroyed and a naked singularity appears , it implies that the test particle has a maximum value $E_{max}$ to destroy the black hole. And Eq.(\ref{afterX}) ensures that Eq.(\ref{aftercaptureNear}) can be used as a physical criterion for destroying the black hole.

Now let us consider a test particle with electric charge $q$, energy $E$, and orbital angular $L$ moves along a geodesic in a Kerr-Newman-AdS spacetime, the conserved energy of the test particle is described as \cite{Siahaan:2015ljs, songyu:2017}
\begin{equation} \label{EtoL}
E=\frac{{g_{t\phi }}}{{g_{\phi\phi}}}\left({qA_\phi-L}\right)-qA_t+{\sqrt{{\left({\frac{{g_{t\phi }^2
-g_{\phi\phi}g_{tt}}}{{g_{\phi\phi }^2}}}\right)\left({\left({L-qA_\phi}\right)^2
+m^2g_{\phi\phi}\left({1+g_{rr}\dot r^2+g_{\theta\theta}\dot \theta^2}\right)}\right)}}}.
\end{equation}
And the minimum energy which allows the test particle to reach the event horizon of the Kerr-Newman-AdS black hole given by\cite{songyu:2017}
\begin{eqnarray}
E_{\rm min} = \frac{{a L\Xi+q Q r_+}}{{a^2 + r_+^2}}\label{Emin},
\end{eqnarray}
So we have the allowed the range of energy for the particle to destroy the black hole as
\begin{equation}
\Delta E = E_{\rm max} - E_{\rm min} >0\,.\label{deltaE}
\end{equation}

\begin{figure}
\begin{center}
\includegraphics[scale=1]{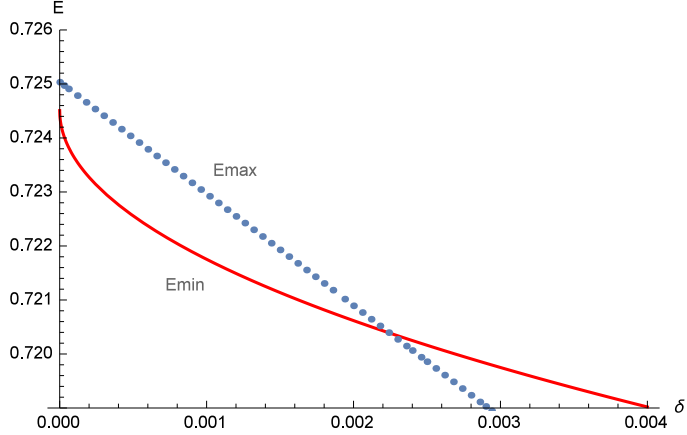}
\end{center}
\caption{Plot of $E$ - $\delta$. The intersection area is  $E < E_{\rm max}$ and $E > E_{\rm min}$.} \label{fig.EmaxEmin}
\end{figure}

To investigate the range of energy $\Delta E$ , we chose the parameters $Q=30$,  $a=0.1$, $q=0.1$, $L=0.1$, then the radius of extremal black hole is $4.1416288$ and mass $M$ is $146.2664986$. By choosing the values of $\delta$, we can get the corresponding values of $E_{max}$ by solving Eq.(\ref{aftercaptureNear}), so one can gets a set of values $[\delta,  E_{max}]$ for plotting the curve of $\delta - E_{max}$, which we provide in FIG.\ref{fig.EmaxEmin}.

In order to get the minimum energy for the particle to reach the event horizon,
We can rewrite Eq.~(\ref{nearextrformalcon}) as
\begin{equation} \label{rewriteextremal}
\sqrt {u-A}(u+A)- \delta \sqrt {u-A}=4 M_{ne} l^2,
\end{equation}
where $A=a^2+l^2$. Solving above equation, we get
\begin{eqnarray}
&&u=\frac{1}{3}(-A-2\delta-\frac{-4A^2-4A\delta-\delta^2}{B}+B)\label{Eminu1},
\end{eqnarray}
where
\begin{eqnarray}
B&=&(8A^3+12A^2\delta+6A \delta^2+\delta^3+216  l^4 M_{ne}^2 \nonumber\\
&&+12\sqrt{3}\sqrt{8A^3l^4 M_{ne}^2+12A^2\delta l^4 M_{ne}^2+6A \delta^2l^4 M_{ne}^2+ \delta^3 l^4 M_{ne}^2+ 108l^8 M_{ne}^4}\,\,)^\frac{1}{3}\nonumber.
\end{eqnarray}\nonumber
consequently, we have the event horizon as
\begin{eqnarray}
r_{+}&=& \frac{1}{2}\left(X+Y \right)\,\,,\label{afterhorizons}
\end{eqnarray}
where
\begin{eqnarray}
X&=&\sqrt{u-A}\,\,,~~~~~ Y= \sqrt{-u -A +\frac{4 M_{ne} l^2}{X}}.\nonumber
\end{eqnarray}

Based on Eqs.~(\ref{Emin}),(\ref{Eminu1}),(\ref{afterhorizons}), By using the same parameters $Q=30$,  $a=0.1$, $q=0.1$, $L=0.1$ as mentioned above, we can plot the curve of $\delta - E_{min}$ which we show in FIG.\ref{fig.EmaxEmin}.

\begin{figure}
\begin{center}
\includegraphics[scale=1]{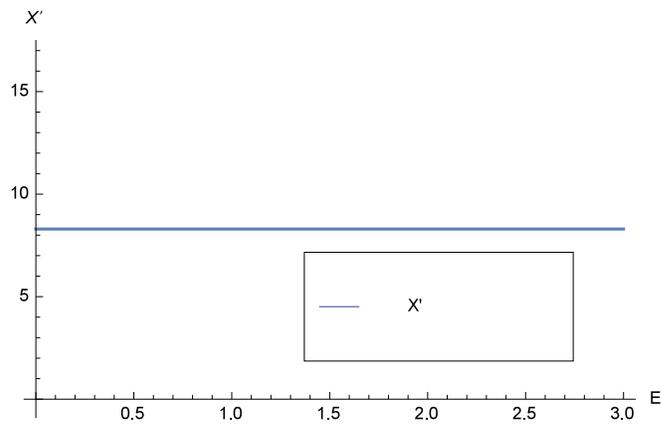}
\end{center}
\caption{Plot of $ E- X' $. $X\geq0$ assure the extremal condition of black hole we used as a judgment is meaningful.} \label{fig.EmaxX}
\end{figure}

To prove Eq.(\ref{afterX}) is satisfied, using the same numerical values with $\delta$ chosen\footnote{From Eq.(\ref{Eminu1}), we can see the value of function $u$ will decrease as $\delta$ increases. So if $\delta$ take values in $[0, 0.004]$, $\delta$ is chosen to be 0.004 will give the minimum value of X'.} to be 0.004 , we provide FIG.\ref{fig.EmaxX}.

 From FIG.\ref{fig.EmaxX}, we can see when particle's energy E range from $0$ to $3$,  the condition Eq.~(\ref{afterX}) is always satisfied.  FIG.\ref{fig.EmaxEmin} shows $\Delta E$ gets narrower as $\delta$ increases, this implies that the black hole is harder to destroy when it moved from the extremality. we can see there is a narrow range of the test particle's energy, $\Delta E$, which allows the appearance of naked singularity from both the near-extremal and extremal Kerr-Newman-AdS black hole. For illustration, we take $\delta=0.001$ and find $0.721753 < E < 0.722962$, so $\Delta E = 0.001209$. for extremal case($\delta = 0$), we get $0.724507 < E < 0.725032$ and $\Delta E = 0.000525$.

\begin{figure}
\begin{center}
\includegraphics[scale=1]{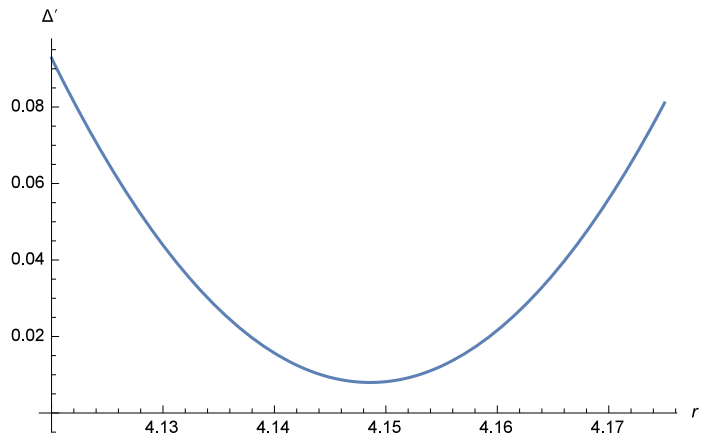}
\end{center}
\caption{Plot of $\Delta{'} - r$.  $E=0.722$ with $\delta$ chosen as 0.001.} \label{fig.destroy}
\end{figure}

To prove the black hole horizon is destroyed, we chose $E=0.722$, which is just between $E_{\rm max}$ and $E_{\rm min}$, The numerical values are the same as those in obtaining FIG.\ref{fig.EmaxEmin}, with $\delta$ chosen as 0.001. We get the plot of $\,\Delta_{\rm }' - r\,$ in FIG.\ref{fig.destroy}.

\begin{figure}
\begin{center}
\includegraphics[scale=1]{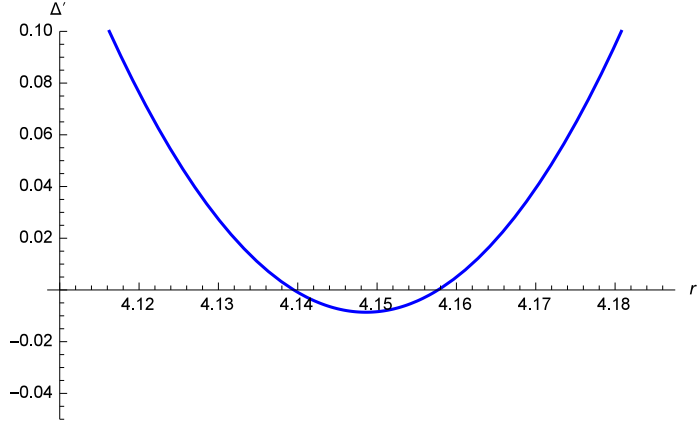}
\end{center}
\caption{Plot of $\Delta{'} - r$. E = $0.730$ with $\delta$ chosen as 0.001.} \label{fig.notdestroy}
\end{figure}

From FIG.\ref{fig.destroy}, we can see the real solution does not exist anymore. This implies that the event horizon is destroyed and the naked singularity is produced after absorbing a test particle. On the other hand, if we chose E is $0.730$,  which is slightly larger than the $E_{\rm max}$, with $\delta$ chosen to be 0.001, we get the plot of $\,\Delta_{\rm }' - r\,$ in FIG.\ref{fig.notdestroy}.

 FIG.\ref{fig.notdestroy} shows there are two different solution of $\,\Delta_{\rm }'\,$. This means after capturing a test particle, the extremal black hole will have inner horizon and outer horizon. Thus it turns into a common black hole. This implies when the energy of test particle located 
 in $[0.721753, 0.722962]$, the black hole will be destroyed.

\begin{figure}[!ht]
\begin{center}
\includegraphics[scale=1]{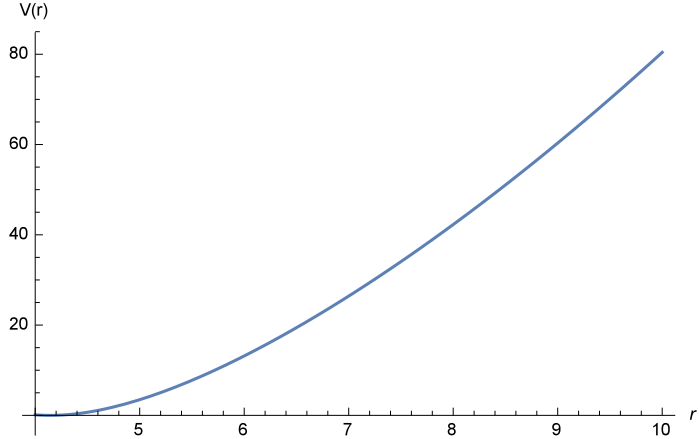}
\end{center}
\caption{The effective potential $V(r)$ of a test particle outside event horizon with $4 <  r  < 10$. test particle's energy $E_{\rm min} <  E  <  E_{\rm max}$.}\label{neareffctive}
\end{figure}

Next, we will show that a particle from infinity whose energy between $E_{\rm max}$ and $E_{\rm min}$ could
really fall into the black hole. Since the metric is axisymmetric, the test particle can fall along the orbit which lying entirely in the plane $\theta=\pi/2$.  For such an orbit, one can solve Eq.~(\ref{EtoL}) for $\dot r^2$ \cite{Gao:2012ca}, and obtain
\begin{equation}
\dot r^2=-V(r),
\end{equation}
where the effective potential $V(r)$ is given by
\begin{eqnarray}
V(r)&=&\frac{1}{l^4 m^2 r^4}[(a^4 L^2-2 a^2 l^2 L^2 +l^4 L^2 +l^4 m^2 r^2)\Delta_r-l^4 E^2 ((a^2+r^2)^2-a^2 \Delta_r)\nonumber\\
&&+2l^2 E ((a^2- l^2) a L \Delta_r -( a^3 L -a l^2 L-l^2 Q q r)(a^2+r^2)\nonumber\\
&&-(a l^2 L-a^3 L +l^2 Q q r)^2],
\end{eqnarray}\label{vvr}
with
\begin{equation}
\Delta_r = \left(r^2 + a^2\right)\left(1 +\frac{r^2}{l^2}\right) -
2 M_{ne} r + Q^2,
\end{equation}
where
\begin{equation}
M_{ne} = \frac{\sqrt{-a^2-l^2+\frac{1}{3}l^2\sqrt{\left(1+\frac{a^2}{l^2}\right)^2+\frac{12
\left(a^2+Q^2\right)}{l^2}}} [a^2+l^2+\delta+\frac{1}{3}l^2 \sqrt{\left(1+\frac{a^2}{l^2}\right)^2+\frac{12
\left(a^2+Q^2\right)}{l^2}}]}{4l^2}.\nonumber
\end{equation}
By choosing the same parameters with $\delta$ chosen to be 0.001, and $m = E = 0.722$ such that $E$ is in the allowed range,
We plot FIG.\ref{neareffctive},  which
shows curve is descending and the particle is easy to fall into a Kerr-Newman-AdS black hole from infinity. The choice $m = E$
indicates that the particle stays at rest relative to a stationary observer at infinity, so it is realizable in practice.

\section{Discussion and Conclusions}\label{sec.2}

In this paper, we have found that a test particle may destroy the horizon of a near-extremal Kerr-Newman-AdS black hole, resulting in an apparent violation of the cosmic censorship without taking into account the radiative and self-force effects. Moreover, the allowed range for the particle's energy $E$ to destroy the black hole is narrow, which indicates that the energy of the particle must be finely tuned. Generally, It is regards that taking the radiative and self-force effects \cite{CardosoPRL,CardosoPRD,Zimmerman:2012zu} into account could compensate for the WCC violation.

It is worth to point out that in \cite{Gao:2012ca,Siahaan:2015ljs,Saa:2011wq}, the higher order term in evaluating the allowed upper limit of particle's energy $E_{max}$ is neglected. While in our paper, we sacrifice the analyticity of $E_{max}$ and use the numerical method , so we can plot the allowed range for test particle's energy to destroy the Kerr-Newman-AdS black hole without approximation. There is a bit queer that the FIG.\ref{fig.EmaxEmin} we obtained is similar to the FIG.4.1 in Ref.\cite{Siahaan:2015ljs},  both of FIG.\ref{fig.EmaxEmin} and FIG.4.1 show $\Delta E$ is very narrow and gets narrower as $\delta$ increases in the near extremal case.

Nevertheless, the analysis performed in this paper neglects the radiative and self-force effects. The fact that $\Delta E$ is very small leaves the possibility of radiative and self-force effects considerations to cure the problem of WCC violation in Kerr-Newman-AdS space time. We will address these projects in our future work.

\section*{Acknowledgements}

This work was supported by the National Science Foundation of China under Grand No. 11275099, 11435006, 11405130.
We are extremely grateful to Cheng-Yi Sun, Hao Tang, Bin Wu, Ming Zhang for useful discussions.

\end{document}